\documentclass[prl,twocolumn,showpacs]{revtex4}
\usepackage{amsmath}
\usepackage{amssymb}
\usepackage{graphicx}
\usepackage{color}

\begin{document}

\title{Hidden Black: Coherent Enhancement of Absorption in
  Strongly-scattering Media}

\author{Y.~D.~Chong}
\email{yidong.chong@yale.edu}

\author{A.~D.~Stone}

\affiliation{Department of Applied Physics, Yale University, New
  Haven, Connecticut 06520}


\begin{abstract}
We show that a weakly absorbing, strongly scattering (white) medium
can be made very strongly absorbing at any frequency within its
strong-scattering bandwidth by optimizing the input electromagnetic
field.  For uniform absorption, results from random matrix theory
imply that the reflectivity of the medium can be suppressed by a
factor $\sim (\ell_a/\ell)N^{-2}$, where $N$ is the number of incident
channels and $\ell,\ell_a$ are the elastic and absorption mean free paths
respectively.  It is thus possible to increase absorption from a few
percent to $> 99\%$. For a localized weak absorber buried in a
non-absorbing scattering medium, we find a large but bounded
enhancement.
\end{abstract}

\maketitle

The absorption properties of a medium at a given frequency depend not
only on the atomic and molecular constituents of the medium, but also
on the coherence properties of the incident radiation.  This was
emphasized by recent work demonstrating the phenomenon of coherent
perfect absorption (CPA), in which a cavity containing a weakly
absorbing medium completely absorbs an appropriately-chosen input
field; this is achieved by choosing the field to be the time-reverse
of the lasing mode which the cavity would emit, if the loss medium
were replaced by a gain medium of equal strength
\cite{cpa_prl,cpa_science}.  Other input fields, differing only in the
phase relationships of the incident waves, are only weakly absorbed,
as are incoherent inputs.  Like a laser, CPA only occurs at discrete
frequencies in a given system.  This raises a question: under what
conditions, if any, is it possible to control absorption over a
continuous and large frequency range by optimizing the input fields at
each frequency?

A related question, which has recently been extensively studied,
concerns the conditions under which light can be transmitted through
an opaque, {\it lossless}, multiple-scattering medium.  In the
diffusive regime, the transmission probability decays as $\ell/L$, where
$\ell$ is the elastic transport mean free path and $L$ is the sample
length; hence a sample with $L \gg l$ is highly reflecting at all
frequencies.  Nonetheless, it has been known for some time that at any
frequency there do exist coherent superpositions of input fields which
penetrate much further than the mean free path, and hence may transmit
with high probability through the sample
\cite{dorokhov84,imry,pendry,beenRMP}.  This effect has been
demonstrated experimentally in the past few years, as practical
methods were found for determining the optimal input fields without
detailed {\it a priori} knowledge of the scattering configuration
\cite{mosk,popoff}.  This suggests that if a multiple-scattering
medium contains some material absorption, then even if that absorption
is weak (i.e., the medium normally appears ``white''), it should be
possible to substantially enhance the \textit{effective} absorption at any frequency
within a wide and continuous frequency range, by tuning the input fields
to penetrate deeply enough that the photon path length exceeds the
large but finite absorption length.  In this Letter, we provide a
theoretical and numerical demonstration of this effect, which we refer
to as {\it coherently enhanced absorption} (CEA). It is in general
distinct from CPA, which follows rigorously from time-reversal
symmetry and is not specific to multiple scattering media (but
requires fine-tuning of the sample parameters and operating
frequency).

\begin{figure}
\centering
\includegraphics[width=0.43\textwidth]{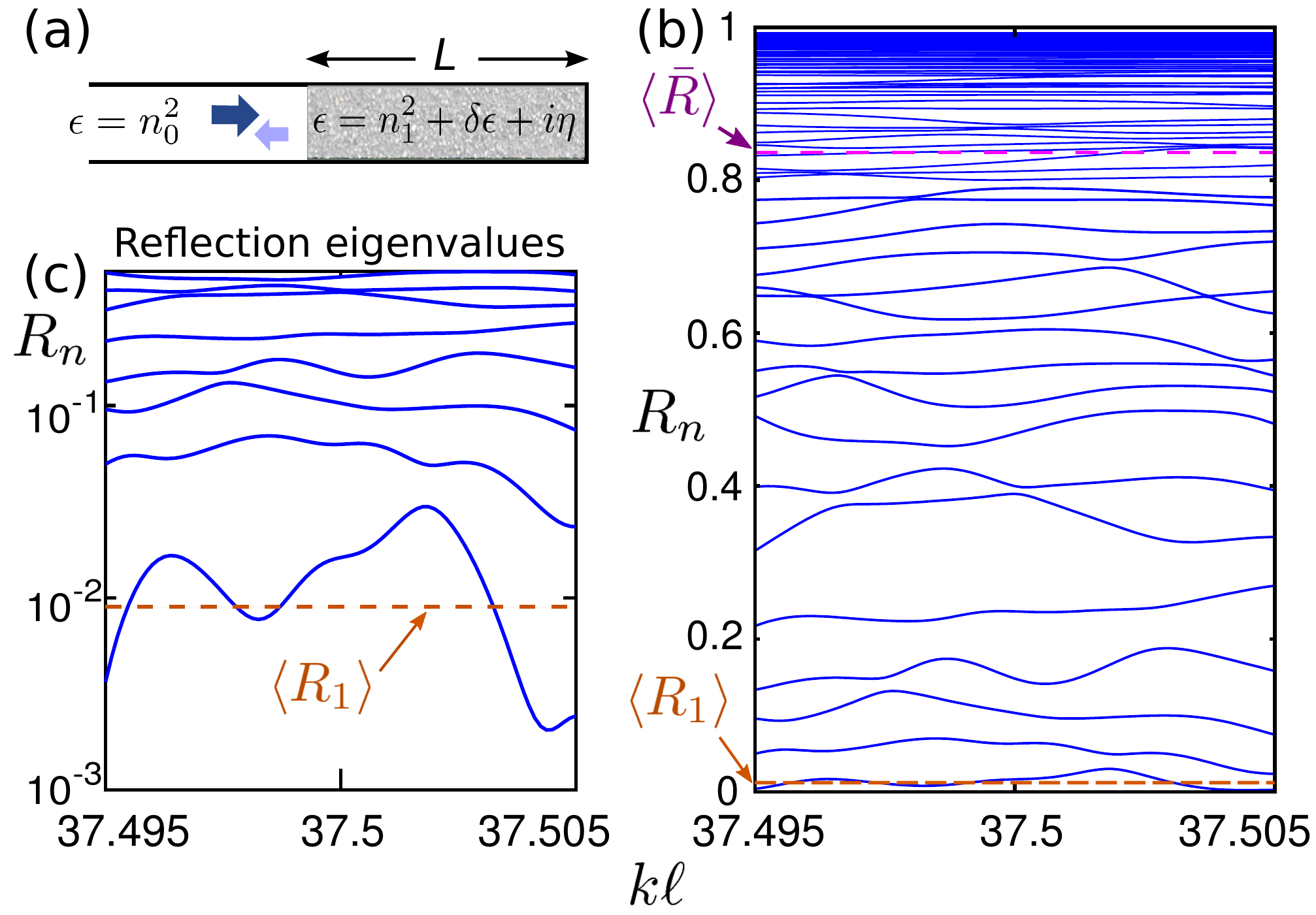}
\caption{(color online) (a) Model of an absorptive multiple-scattering
  system.  A waveguide with dielectric constant $n_0^2$ is connected
  to a region of length $L$ with dielectric $n_1^2 + \delta\epsilon +
  i \eta$, where $\delta\epsilon (x) $ is a real white-noise disorder
  term drawn from a uniform distribution over $[-d_0,d_0]$, and $\eta$
  is the material absorption.  (b) Variation of reflection eigenvalues
  $\{R_n\}$ with frequency $k$ for a {\it fixed} disorder realization, with
  $n_0 = n_1 = 1.5$, $N=80$ scattering channels, mean free path $\ell
  = 0.05L$, and uniform $\eta$ with ballistic absorption length
  $\ell_a = 13.3L$.  Dashed lines show the ensemble averages of the
  mean and smallest reflectivity, $\bar{R}$ and $R_1$.  (c) Semi-log
  plot of $R_1$, which fluctuates, but is $\sim 10^{-2}$ over a continuous frequency range.}
\label{kplot}
\end{figure}

Linear scattering from a medium is described by the scattering matrix
(S-matrix), which relates incident and scattered amplitudes in the
basis of asymptotic free solutions of the wave equation.  The S-matrix
depends on the scattering geometry and material properties, as well as
the frequency of the input fields, but does not depend on the fields
themselves.  Given a vector $|\psi_\textrm{in}\rangle$ whose
components are the intensity-normalized input amplitudes in each of
the $N$ scattering channels, the output amplitudes are
$|\psi_\textrm{out}\rangle = S |\psi_\textrm{in}\rangle$, and the
total output intensity is $\left\langle \psi_\textrm{out}
|\psi_\textrm{out}\right\rangle = \langle \psi_\textrm{in} |S^\dagger
S|\psi_\textrm{in}\rangle$.  For a lossless system, this quantity is
always unity since $S$ is unitary; in the presence of absorption, it
is smaller than one \cite{beenakker0}.

We will focus on the case of scattering from one surface of an
optically dense and weakly absorbing medium, in which all incident
fields are either reflected or absorbed, so that the S-matrix
coincides with the reflection matrix $r$.  Denote the eigenvalues of
the Hermitian matrix $r^\dagger r$ by $R_n$, where $0 \le R_1 \le
\cdots \le R_N < 1$.  The sample-specific ``mean reflectivity'' $\bar
R = (1/N) \sum_n R_n$ is the normalized output power when the inputs
in each channel are independent (e.g.~having no coherent phase
relationship); the moments of this quantity were studied in earlier
works on reflection from random absorbing media
\cite{beenakker,BruceChalker}.  In studying CEA, the relevant physical
quantity is the \textit{smallest} reflection eigenvalue, $R_1$.  By
the variational principle, $\langle \psi_\textrm{in} |r^\dagger r
|\psi_\textrm{in}\rangle \ge R_1$ for any intensity-normalized
$|\psi_\textrm{in}\rangle$.  Hence, $1-R_1$ is the maximum amount of
absorption that can be achieved, and its eigenvector gives the
maximally-absorbed set of input amplitudes.  In an experiment, if
noise sources rule out measuring reflectivities below a value $R' >
R_n$, any input waveform in the eigenspace spanned by eigenvectors of
$R_1,R_2 \ldots R_n$ will give reflectivity $\sim R'$.  However, we
note that enhanced absorption can also be measured directly, e.g.~via
fluorescence or heat generation.

We employ a model previously introduced and studied by Beenakker
\textit{et.~al.}  \cite{beenakker}, and by Bruce and Chalker
\cite{BruceChalker}.  As shown in Fig.~\ref{kplot}(a), it consists of
a waveguide with $N$ channels, containing a medium with elastic
scattering mean free path $\ell$ and ballistic absorption length
$\ell_a$.  The ratio of these lengths defines the absorption parameter
$a \equiv \ell / \ell_a $, with $a \ll 1$ in the regime of interest.
In addition, the quantity $L_a \equiv \sqrt{\ell_a \ell}$ gives the
average distance a diffusing particle penetrates into the medium
before being absorbed.  From the combination law for infinitesimal
scattering segments (each of which randomly scatters between channels
and also imposes some absorption parameterized by $a$), the joint
probability distribution (JPD) for the reflection eigenvalues can be
shown to evolve with the sample length $L$ according to the equation
\cite{beenakker,BruceChalker}
\begin{equation}
  \ell (N+1) \frac{\partial p_x}{\partial L} = \frac{1}{2} \sum_n
  \frac{\partial^2 p_x}{\partial x_n^2} + \frac{1}{2} \sum_n
  \frac{\partial}{\partial x_n} \left( \frac{\partial V}{\partial x_n}
  p_x\right),
  \label{DMPK}
\end{equation}
where $p_x(\{x_n\})$ is the JPD given in terms of the variables $x_n
\equiv \frac{1}{2} \cosh^{-1}[(1+R_n)/(1-R_n)]$.  This change of
variables leaves the diffusion term in (\ref{DMPK}) independent of
$\{x_n\}$, allowing analytic solution of the stationary distribution.
The ``potential" $V$ in Eq. (\ref{DMPK}) is given by
\begin{multline}
  V = \sum_n \Big[a(N+1) \cosh(2x_n) - \ln|\sinh(2x_n)|\Big] \\ -
  \sum_{m>n} \ln\,\left[\cosh(2x_m) - \cosh(2x_n)\right].
  \label{DMPK potential}
\end{multline}

Eq.~(\ref{DMPK}) is similar to the Dorokhov-Mello-Pereyra-Kumar (DMPK)
diffusion equation \cite{DMPK1,DMPK2}, which is derived from a
quasi-one-dimensional (quasi-1D) waveguide model without absorption.
The DMPK equation describes the JPD of transmission and reflection
eigenvalues, and its results have been found to generalize
to geometries with no transverse confinement, as
when a surface is illuminated by a spot \cite{nazarov94} (the geometry
used in subsequent experiments on coherently enhanced transmission
\cite{mosk}).  We expect the results for the absorbing case to
generalize in the same way.

For $L_a < L$, $p_x(x_1,\cdots,x_N) \sim e^{-V}$ is the stationary
($L$-independent) limiting solution to (\ref{DMPK}).  To make contact
with standard random matrix theory (RMT), a further change of
variables $y_n \equiv 2a(N+1) [\cosh(2x_n) - 1]$ is performed, which
yields
\begin{equation}
  p_y(y_1,\cdots,y_N) \sim e^{-\frac{1}{2}\sum_n y_n} \prod_{m>n} (y_m
  - y_n).
  \label{stationary DMPK solution}
\end{equation}
This is the Laguerre eigenvalue distribution, characteristic of a
well-studied ensemble in RMT \cite{Edelman}.  The ensemble-averaged
mean reflectivity can be determined by integrating over this
distribution \cite{beenakker,BruceChalker}:
\begin{equation}
  \langle \bar{R} \rangle \approx 1 + 2a - 2 \sqrt{a(1+a)}.
  \label{sigmabar}
\end{equation}
Note that this result is independent of $N$.

We are interested in the distribution of the smallest reflection
eigenvalue, $R_1$.  This can be found from the fact that $Ny_1$
follows a $\chi^2_2$ distribution \cite{Edelman}.  Thus,
\begin{equation}
  p(R_1) = 2aN(N+1)
  \frac{e^{-2aN(N+1)\frac{R_1}{1-R_1}}}{(1-R_1)^2},
  \label{psigma_dist}
\end{equation}
and the ensemble-averaged value is
\begin{eqnarray}
  \langle R_1\rangle &=& 1 + 2aN(N+1) e^{2aN(N+1)}\,
  \textrm{Ei}[-2aN(N+1)] \nonumber \\ &\approx& \left[2aN(N+1)\right]^{-1}
  \;\;\textrm{for}\;\; a \gtrsim N^{-3/2}.
  \label{sigma1}
\end{eqnarray}
Unlike $\langle \bar{R}\rangle$, this quantity $\textit{does}$ depend
on $N$.  The decrease of $R_1$ with increasing $N$ is very natural,
because the eigenvalue repulsion characteristic of RMT ensembles tends
to push the smallest eigenvalue to small values.

Thus, even if the material absorptivity is weak enough that
$\langle\bar{R}\rangle \sim 1$, if $N$ is sufficiently large then $R_1
\sim 1/aN^2$ can be very small.  In typical free-space optical
experiments $N \approx A/\lambda^2$ where $A$ is the spot size, so
this condition should be easily achieved.  According to
(\ref{sigmabar}) and (\ref{sigma1}), the regime of weak average
absorption and strong optimal absorption is $a \ll 1 \ll 2aN^2$.  The
physical intepretation of these inequalities is that photons have
little chance of being absorbed between scattering events $(\ell_a \gg
\ell)$, but those in the maximally-absorbed mode have a negligible
chance of diffusing the entire quasi-1D localization length $\xi = N
\ell$ before being absorbed.  For example, for $a \leq 0.01$ and $N =
80$, we have $\langle\bar{R}\rangle \geq 0.8$ and $\langle \bar{R}_1
\rangle < 0.01$.

To test and generalize the above results, we have performed numerical
simulations of scalar wave propagation in quasi-1D structures.  The
wave equation $(\nabla^2 + \epsilon k^2) \psi = 0$ is discretized
using a 2D tight-binding model on a square grid. A semi-infinite
uniform waveguide of index $n_0$ is connected to a scattering region
with average index $n_1$, uniform absorption $\eta$ and random
scatterers; see Fig.~\ref{kplot}(a). The reflection matrix is computed
for this rectangular region, of varying length and width and perfectly
reflecting walls on three sides, using the recursive Green's function
method \cite{beenakker,RGF}.  The elastic mean free path $\ell$ is
computed using the relation $\langle \bar{T} \rangle \equiv \ell/L$,
via separate transmission simulations without absorption.  In all
simulations, we use frequency $k = 750/L$, lattice spacing $L/750$,
and waveguide width $L/5$. The ballistic absorption length is
$\ell_a^{-1} = c_0 \eta k/2 n_1$, where $c_0$ is a constant of order
unity determined by a fit to Eq.~(\ref{sigmabar}).

\begin{figure}
\centering
\includegraphics[width=0.45\textwidth]{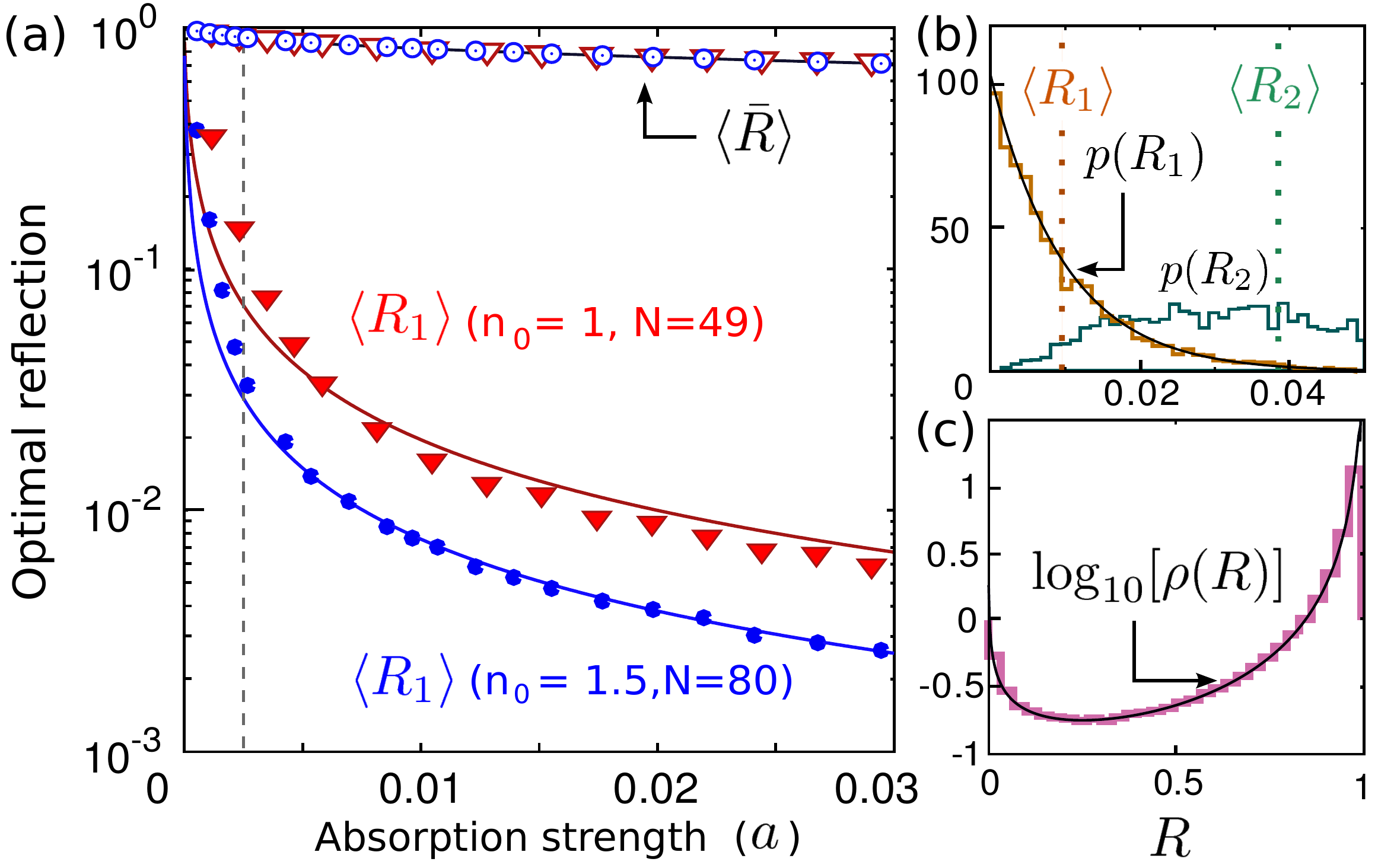}
\caption{(color online) (a) Mean reflectivity $\bar{R}$ (open symbols)
  and smallest reflection eigenvalue $R_1$ (filled symbols)
  \textit{vs} the absorption strength $a$.  Each data point is an
  average over 2000 disorder realizations.  The scattering medium has
  $n_1 = 1.5$, disorder strength $d_0 = 0.6$ ($\ell = 0.05L$), and
  uniform absorption.  Two simulation sets are shown: (i) external
  index $n_0 = 1.5$ and $N=80$ exterior waveguide channels (circles);
  and (ii) index mismatch with $n_0 = 1$ and $N=49$ exterior channels
  (triangles).  In each case, we fit $\langle\bar R\rangle$ to
  Eq.~(\ref{sigmabar}) to obtain the constant $c_0$ in $\ell_a^{-1} = c_0
  \eta k/2 n_1$, finding $c_0 = 2.14$ for (i) and $c_0 = 4.65$ for
  (ii).  This determines the analytic form of $\langle R_1\rangle$,
  plotted as solid curves, with no free parameters.  The vertical
  dashed line denotes $a \geq (\ell/L)^2$, required for the stationary
  solution to apply.  (b) Distributions of $R_1$ and $R_2$ from
  simulations (histograms).  The solid line shows the analytical
  expression (\ref{psigma_dist}) for $p(R_1)$; dotted lines show the
  ensemble averages $\langle R_1\rangle$ and $\langle R_2\rangle$.
  (c) Semi-log plot of reflection eigenvalue density, comparing
  simulations (histogram) to the approximate density
  $2a(N+1)\sqrt{[N(1-R)/a(N+1)R]-1}/\pi(1-R)^2$ (solid line), from the
  large-$N$ density of the Laguerre distribution \cite{Edelman}.  The
  simulations for (b) and (c) were performed with $n_0 = 1.5$ ($N =
  80$) and $\eta = 0.0003$ ($\ell_a = 125\ell$).}
\label{sigmaplot}
\end{figure}

The variation of $\langle \bar R \rangle$ and $\langle R_1\rangle$
with $\ell/\ell_a$ is shown in Fig.~\ref{sigmaplot}(a), together with
the analytic predictions.  Two sets of simulations are plotted.  In
the first, the lead and scattering medium are index-matched ($n_0 =
n_1 = 1.5$) as in the model of \cite{beenakker,BruceChalker}.  In the
second, they are mismatched with $n_0 = 1$, $n_1 = 1.5$, representing
scattering from air into a denser medium, a case for which analytic
results were not known.  In both cases, despite weak average
absorption ($\ell_a > 40 \ell$ and $\langle \bar R \rangle \sim 1$),
we obtain values of $\langle R_1\rangle$ as small as $10^{-2}$ to
$10^{-3}$.  For the index-matched results the data for $\langle R_1
\rangle$ agree very well with the theory, Eq.~(\ref{sigma1}), using
the number of exterior channels, $N=80$.  For the index-mismatched
case, the exterior waveguide has fewer channels, $N=49$, at the same
frequency, compared to the average of 80 interior channels;
nonetheless, we find that Eq.~(\ref{sigma1}) holds very well using the
exterior channel number.  Thus, illuminating from a lower index medium
(e.g. from air) does diminish the CEA effect compared to the
index-matched case, but large enhancement is still possible and is
governed by the same factor of $(2aN^2)^{-1}$. 
Figs.~\ref{sigmaplot}(b) and (c) show the numerical distributions of
the reflection eigenvalues for fixed $a$, which are in excellent
agreement with the theoretical distributions.  The distribution of
$R_2$ in Fig.~\ref{sigmaplot}(b) illustrates the role of eigenvalue
repulsion in confining $R_1$ to small values.

\begin{figure}
\centering
\includegraphics[width=0.35\textwidth]{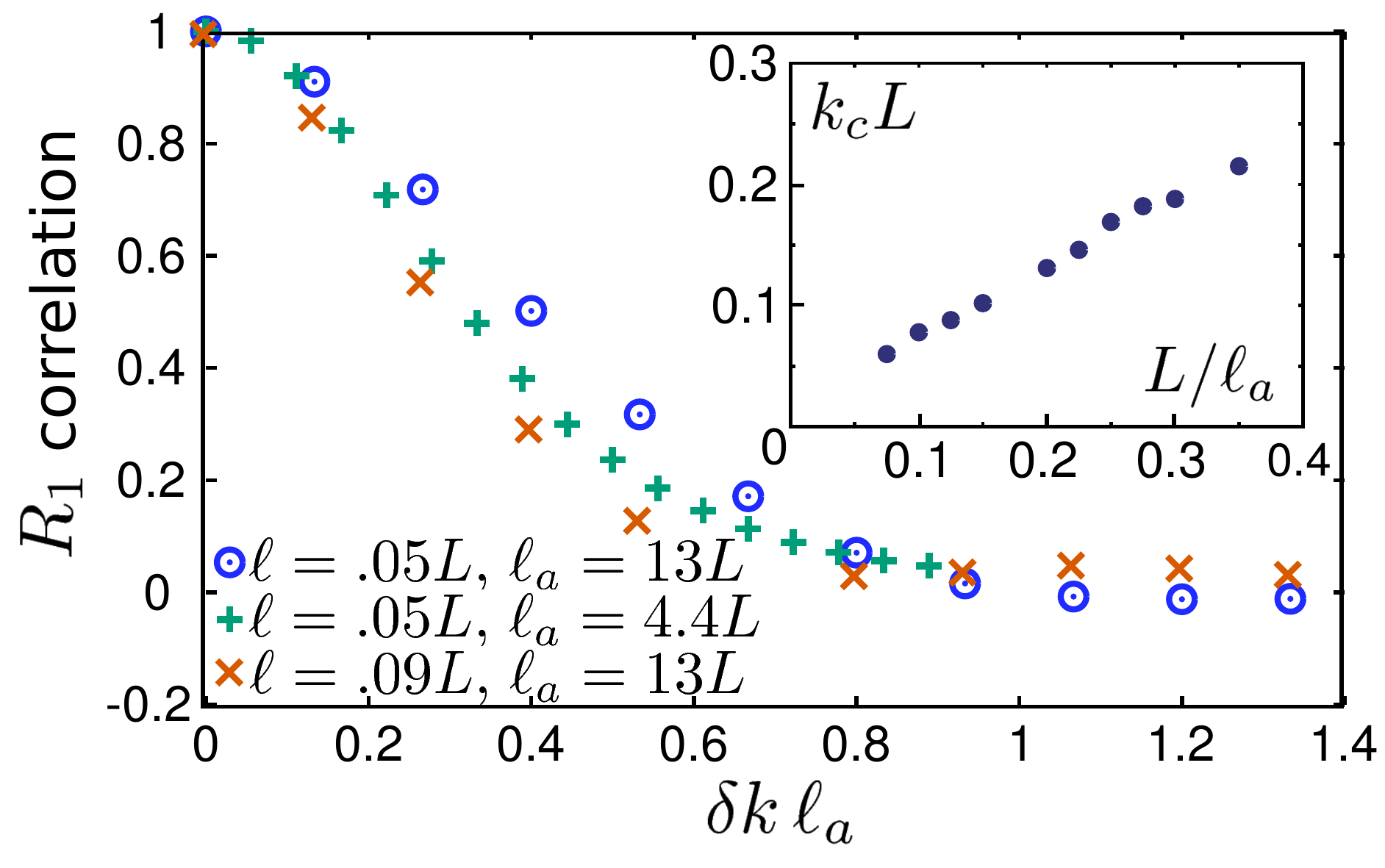}
\caption{(color online) Normalized correlation function for the
  smallest reflection eigenvalue $R_1$, defined as $C_{R_1}(\delta k)
  \propto [\langle R_1(k+\delta k) R_1(k)\rangle - \langle
    R_1(k)\rangle^2]$ with $C_{R_1}(0) = 1$.  Each data set consists
  of 3000 disorder realizations.  Inset: plot showing the linear
  dependence on $1/\ell_a$ of the decorrelation frequency $k_c$,
  defined by $k_c \equiv C_{R_1}(k_c/2)=1/2$.  Here, $\ell = 0.05L$.}
\label{kcorplot}
\end{figure}

The analytic predictions for $\langle R_1 \rangle$ and $\langle \bar R
\rangle$ have no dependence on the frequency $k$, apart from the weak
variation of the number of exterior channels $N$ with $k$, assuming
that the parameter $a$ varies negligibly with frequency.  Unlike the
CPA systems studied in Refs.~\cite{cpa_prl,cpa_science}, in which
near-perfect absorption is obtained at particular discrete values of
$a$ and $k$, in the present systems highly enhanced absorption is
possible at any frequency {\it so long as the incident waveform is
  optimized in the vicinity of each frequency of interest.}  Thus,
while CEA is not a broadband effect in the usual sense, it can be
realized over a wide range of frequencies.  To determine the frequency
interval over which a fixed waveform, optimized at a single frequency,
will still lead to strongly enhanced absorption, we calculate the
frequency correlation of $R_1$, shown in Fig.~\ref{kcorplot}.  We find
that the decorrelation frequency scale is $k_c \approx 0.5 /\ell_a$,
which is to be expected; by analogy to the case of transmission
through a non-absorbing medium, where the decorrelation scale is
determined by the length of paths that traverse the sample and escape,
here the maximally-absorbed mode arises from the interference of many
coherent paths of length $\sim \ell_a$.  Thus, measurements obtained
at frequency intervals $\Delta k \gg 0.5/\ell_a$ are independent.  For
$ \ell = 1 \mu \rm{m}$, $\ell_a = 100 \ell$, and $\lambda = 1.5 \mu
\rm{m}$, this gives $\Delta \lambda_c \approx 1.5 \,{\rm nm}$.  While
the CEA effect is more robust than CPA when $N \gg 1$, it should be
noted that, when $N \to 1$, continuous CEA is no longer possible,
whereas CPA can still occur at discrete frequencies.

\begin{figure}
  \centering
  \includegraphics[width=0.423\textwidth]{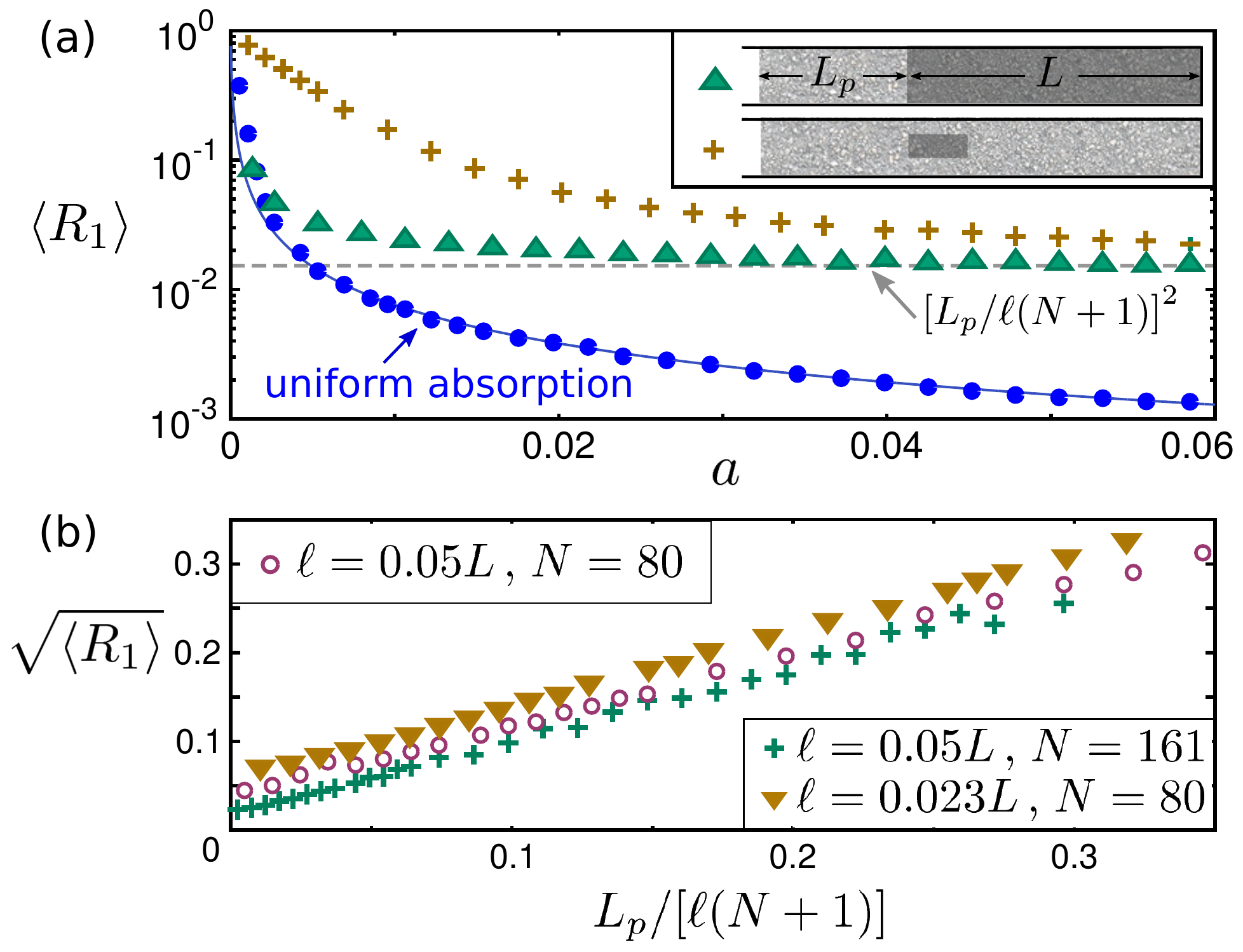}
  \caption{(color online) (a) Variation of $R_1$ with $a$ for buried
    absorbers.  In the first set of simulations (triangles), the
    absorbing region (length $L$, width $0.2L$ equal to the waveguide
    width, and $\ell = 0.05L$) lies behind a segment of lossless
    scattering material (length $L_p = 10\ell$, with the same $\ell$).
    In the second set of simulations (crosses), the absorbing region
    has dimensions $0.06L \times 0.08 L$ and is surrounded by the
    lossless medium.  The uniform absorption case, similar to
    Fig.~\ref{sigmaplot}, is shown for comparison (circles).  (b) Plot
    of $\sqrt{\langle R_1\rangle}$ \textit{vs} the Fokker-Planck
    ``time'' $L_p/\ell(N+1)$, for the full-width absorber with $a =
    0.04$.  Note that $R_n \approx x_n^2$, where $x_n$ are the
    variables of Eq.~(\ref{DMPK}).}
  \label{Lpplot}
\end{figure}

An interesting and important extension of the above model is to a
localized or extended absorber buried behind a layer of lossless
scattering medium.  Such a system has been studied experimentally by
Vellekoop \textit{et.~al.}, who demonstrated strong focusing of light
through a lossless medium to enhance fluorescence in a small interior
spot \cite{fluor}.  To our knowledge, there has been no theoretical
work on the limiting efficiency of such a process.  We consider two
variant models, shown in the inset of Fig.~\ref{Lpplot}(a).  A
lossless region of length $L_p$ is added in front of the absorbing
region of length $L$, which contains either uniform absorption or a
localized absorber.  For $a$ within the absorber satisfying $2aN^2 \gg
1$, the simulations still show CEA of approximately two orders of
magnitude.  However, $R_1$ now saturates with increasing $a$ to a
value $\approx (L_p/\ell N)^2$.  This saturation value appears to be
independent of the size of the absorbing region, though the larger
region reaches this value for smaller $a$.

To understand this behavior qualitatively, we return to
Eq.~(\ref{DMPK})-(\ref{DMPK potential}).  One can think of the
eigenvalues $\{x_n\}$ as interacting gas particles with a stationary
distribution at length $L$, in which the confining potential due to
$a$ in (\ref{DMPK potential}) is balanced by diffusion and
interparticle interactions \cite{beenRMP}.  This is the distribution for which
$\langle R_1 \rangle \approx (2aN^2)^{-1}$ for the extremal particle.
Adding the lossless region corresponds to turning off the confining
potential for a ``time'' $T = L_p/\ell (N+1)$, causing the particles
to drift and diffuse to larger values of $x_n$.  In fact,
Fig.~\ref{Lpplot}(b) shows that $\langle x_1(L_p)\rangle \sim T$, so
that $\langle R_1 \rangle \approx \langle x_1^2 \rangle \sim T^2$,
which is the saturation scale observed in Fig.~\ref{Lpplot}(a).  For
$L_p \lesssim \ell(N+1)$, these results show the feasibility of
enhancing the delivery of energy to a buried absorber by orders of
magnitude.

A final critical question is whether incident waveforms approaching
the CEA optimum can be found when details of the scattering medium are
unknown, via an empirical optimization scheme analogous to those
employed for transmission through lossless media \cite{mosk,popoff}.
Initial numerical tests based on our model are promising, with simple
optimization schemes giving order of magnitude increases in
absorption, though not yet approaching theoretical limits.  This
problem, and that of the buried absorber, are rich and exciting
directions for future work.

This work was partially supported by NSF Grant No. DMR-0908437, by seed
funding from the Yale NSF-MRSEC (DMR-0520495), and by the facilities
and staff of the Yale University Faculty of Arts and Sciences High
Performance Computing Center.  We thank Hui Cao and Wenjie Wan for
helpful discussions.

\end{document}